\documentclass[aps,prl,reprint,groupedaddress,nofootinbib]{revtex4-1}

\usepackage{graphicx}
\usepackage{longtable}

\usepackage{cancel}
 \usepackage[normalem]{ulem}
\usepackage{color}

\def\lapp{\mathrel{\rlap{\raise.5ex\hbox{$<$}}
                    {\lower.5ex\hbox{$\sim$}}}}
\def\gapp{\mathrel{\rlap{\raise.5ex\hbox{$>$}}
                    {\lower.5ex\hbox{$\sim$}}}}

\begin{document}


\title{Bounds on Universal Extra Dimension from LHC Run I and II data}


\author{\vskip -10pt Debajyoti Choudhury$^1$ and Kirtiman Ghosh$^2$}
\affiliation{Department of Physics and Astrophysics, University of Delhi, Delhi 110007, India\\
$^1$debajyoti.choudhury@gmail.com, $^2$kirti.gh@gmail.com}




\begin{abstract}
We discuss the collider bounds on minimal Universal Extra Dimension
(mUED) model from LHC Run-I and II data. The phenomenology of mUED is
determined by only two parameters namely, the compactification scale
($R^{-1}$) of the extra dimension and cutoff scale ($\Lambda$) of the
theory. The characteristic feature of mUED is the occurrence of nearly
degenerate mass spectrum for the Kaluza-Klein (KK) particles and
hence, soft leptons, soft jets at the collider experiments. The degree
of degeneracy of KK-mass spectrum crucially depends on $\Lambda$.  The
strongest direct bound on $R^{-1}$ ($\sim $950 GeV for large
$\Lambda$) arises from a search for a pair of soft dimuons at the
Large Hadron Collider (LHC) experiment with 8 TeV center-of-mass
energy and $20~{\rm fb}^{-1}$ integrated luminosity. However, for
small $\Lambda$ and hence, small splitting within the first KK-level,
the bounds from the dimuon channel is rather weak.  On the other hand,
the discovery of 126 GeV Higgs boson demands small $\Lambda$ to
prevent the scalar potential form being unbounded from below. We
discuss LHC monojet searches as a probe of low $\Lambda$ region of
mUED parameter space. We also compute bounds on the mUED parameter
space from 13 TeV multijets results.
\end{abstract}
\maketitle


%

Theories with one or more extra space-like dimension(s) accessible to
all or a few of the
 Standard Model (SM) fields are of interest for various
reasons. For example, the ADD \cite{add,antoniadis1} (seemingly)
and RS \cite{rs} models provide 
solutions to the long-standing naturalness/hierarchy problem by
postulating the existence of compactified extra-dimension(s)
accessible only to gravity with the SM fields being confined to a
3-brane embedded in the extra-dimensional bulk. On the other hand,
there are a class of models wherein some or all of the SM fields can
access the extended space-time manifold \cite{antoniadis1,acd}, whether fully or
partially. Such extra-dimensional scenarios could lead to a new
mechanism of supersymmetry breaking \cite{antoniadis1}, relax the
upper limit of the lightest supersymmetric neutral Higgs
mass\cite{relax}, give a different perspective to the issue of fermion
mass hierarchy \cite{Arkani-Hamed:1999dc}, interpret the Higgs as a
quark composite leading to a electroweak symmetry breaking (EWSB)
without a fundamental scalar or Yukawa interactions
\cite{Arkani-Hamed:2000hv}, lower the unification scale down to a few
TeVs \cite{dienes}, provide a cosmologically viable candidate for dark
matter \cite{darkued1,darkued2}, explain the long life time of proton
\cite{2UED_proton}, predict the number of fermion generations to be an
integral multiple of three \cite{2UED_fg} and give rise to interesting
signatures at collider experiments. As a result, search for the extra
dimension(s) is one of the prime goals of the Large Hadron Collider
(LHC) experiment \cite{ED_LHC,ATLAS_UED}. Our concern here is a
specific and particularly interesting framework, called the Universal
Extra Dimension (UED) scenario.

The minimal version of UED (mUED) is characterized by a single flat
extra dimension ($y$), compactified on an $S^1 /Z_2$ orbifold with
radius $R$, which is accessed by all the SM particles \cite{acd}.
While a resolution of the hierarchy problem requires that $R^{-1} \sim
{\cal O}(1\, {\rm TeV})$, it has long been argued that, in the absence
of a dynamical stabilization of $R^{-1}$ (or the cutoff), this is just
a postponement of the explanation. However, recently, a mechanism for
the stabilization of $R^{-1}$ has been proposed in the context of
higher-dimensional theories.  The particle spectrum of mUED contains
infinite towers of Kaluza-Klein (KK) modes (identified by an integer
$n$, called the KK-number) for each of the SM fields with the zero
modes being identified as the corresponding SM particles. The key
feature of the UED Lagrangian is the conservation of the momentum
along fifth direction. From a 4-dimensional perspective, this implies
conservation of the KK-number. However, the additional $Z_2$ symmetry
($y\leftrightarrow -y$), which is required to obtain chiral structure
of the SM fermions, breaks the translational invariance along the 5th
dimension. As a result, KK-number conservation breaks down at
loop-level, leaving behind only a conserved KK-parity, defined as
$(-1)^n$. There are several interesting consequences of this discrete
symmetry which, in turn, is an automatic outcome of the $S^1 /Z_2$
orbifolding. KK-parity ensures the stability of the lightest
KK-particle (LKP), allows only pair production of level-1 KK-particles
at the collider, and prohibits KK-modes from affecting tree-level EW
precision observables.  And, although KK-modes do contribute to
standard electroweak processes at higher orders, KK-parity ensures
that, in a loop, they appear only in pairs resulting in a substantial
suppression of such contributions.

Being a higher dimensional theory, mUED is nonrenormalizable and
should be treated as an effective theory valid upto a cutoff scale
$\Lambda$, expected to be somewhat larger than $R^{-1}$.  With
KK-parity ensuring that one-loop\footnote{The observables start
  showing cutoff sensitivity of various degrees as one goes beyond
  one-loop or considers more than one extra dimension.} mUED
corrections to all electroweak observables are cutoff
independent \cite{db}, the latter serve to constrain $R^{-1}$, almost
independent of $\Lambda$.  For example, low energy observables like
muon $g-2$ \cite{g_muon}, flavour changing neutral currents
\cite{chk}, $Z \to b\bar{b}$ decay \cite{santa}, the $\rho$-parameter
\cite{rho}, $\bar{B} \to X_s \gamma$
\cite{Haisch:2007vb} and other electroweak precision tests
\cite{ewued} put a lower bound of about 300-600 GeV on $R^{-1}$.
This, along with the fact that the tree level mUED masses for
level-$n$ KK-excitations are given by $m_n^2 = m_0^2 + n^2 \,R^{-2}$
($m_0$ being the mass associated with the corresponding SM field)
implies that, within a given level, the excitations are quite
degenerate. Quantum corrections partially lift this degeneracy
\cite{radiative} and, typically, $B_1$, the level-1 excitation of
hypercharge gauge boson\footnote{The notation may seem confusing, but
  note that, owing to the the large difference between the electroweak
  scale and $R^{-1}$, the analogues of the Weinberg angle are small
  for the KK-sectors.}, is the LKP, and, hence, stable.  Being only
weakly interacting, the $B_1$ turns out to be a good dark matter
(DM) candidate~\cite{darkued1}.  Consistency with
WMAP/PLANCK--measured \cite{wmap} DM relic density data puts an upper
bound of 1400 GeV on $R^{-1}$. Given this upper limit, it is extremely
plausible that experiments at the LHC can either discover or rule out
mUED. In this paper, we have discussed the impact of LHC Run I
(center-of-mass energy $\sqrt s = 8$ TeV, integrated luminosity ${\cal
  L}=20.3$ fb$^{-1}$) and Run II ($\sqrt s=13$ TeV and ${\cal L}=3.2$
fb$^{-1}$) results on mUED parameter space. In particular, we have
obtained bounds on mUED parameter space from collider upper limits on
the product of cross section, acceptance and efficiency ($\sigma
\times A \times \epsilon$) in monojet \cite{CMS8_monojet,ATLAS8_monojet} and
multijets \cite{ATLAS13_multijets,CMS13_multijets} plus missing energy
($\cancel E_T$) channels.

Given $R^{-1}$ and, hence, the average KK-mass,
  the collider phenomenology is uniquely determined by the mass
  splittings, {\em i.e.}, the quantum corrections.  Apart from the
usual radiative corrections that we expect in a Minkowski-space field
theory, there are additional corrections accruing from the fact of the
fifth direction being compactified on $S_1/Z_2$-orbifold. The
correction terms can be finite (bulk correction) or logarithmically
divergent (boundary correction). Bulk corrections arise only for the
gauge boson KK-excitations due to the winding of the internal loop
(lines) around the compactified direction \cite{radiative} . The 
ubiquitous boundary
corrections are just the counterterms of the total orbifold
corrections, with the finite parts being completely unknown, dependent
as they are on the details of the ultraviolet completion.  Assuming
that the boundary kinetic terms vanish at the cutoff scale $\Lambda$,
the corrections from the boundary terms, at a renormalization scale
$\mu$ would obviously be proportional to $\ln (\Lambda^2
/\mu^2)$. Finite bulk corrections being subdominant, the cutoff scale
plays the most crucial role in determining the
mass-splitting and hence, the collider signatures of level-1
KK-particles. 
The perturbativity of the $U(1)$ gauge coupling
requires that $\Lambda \lapp 40 R^{-1}$. 
It has been argued that a much stronger bound arises from the 
the running of the Higgs-boson self-coupling and the stability of the 
electroweak vacuum \cite{UED_VS2}. However, note that such arguments were based only on 
the lowest-order calculations and the inclusion of higher-loops (which
are poorly understood in this theory) can substantially change these results.
Consequently, we will not impose the last-mentioned.
 In order to discuss the
collider signatures and present the numerical results, we have chosen
three benchmark points (BPs) listed in Table~\ref{BP} along with the
masses of relevant level-1 KK-particles. While the relatively small value of $\Lambda R$ for  BP1 and BP2
is reflected in the approximate degeneracy of level-1 KK-particles,
a much larger value of the same for BP3 results in a wider splitting.

 \begin{table}
 \caption{Benchmark points and mass spectrum of relevant level-1 particles. Total cross-sections ($\sigma^{tot}$) of KK-squarks/gluons pair production at the LHC with 8 and 13 TeV center-of-mass energy are also presented.}
\label{BP}
 \begin{ruledtabular}
\begin{tabular}{||c||c|c||c|c|c|c|c|c||}
BPs & $R^{-1}$ & $\Lambda R$ & $m_{g_1}$ & $m_{Q_1}$ & $m_{q_1}$& $m_{W_1/Z_1}$& $m_{L_1}$& $m_{B_1}$ \\
    & [GeV]& & [GeV]& [GeV]& [GeV]& [GeV]& [GeV]& [GeV]  \\\hline\hline
BP1 & 920 & 3 & 1002 & 973 & 966 & 941 & 929 & 920 \\
BP2 & 1100 & 3 & 1196 & 1163 & 1153 & 1125 & 1111 & 1099 \\
BP3 & 1120 & 35 & 1414 & 1328 & 1300 & 1194 & 1157 & 1119 \\\hline\hline\hline
\multicolumn{9}{||c||}{Total strong pair production cross-section [pb]}\\\hline\hline
\multicolumn{3}{||c||}{$\sqrt s$} & \multicolumn{2}{|c|}{BP1} & \multicolumn{2}{|c|}{BP2} & \multicolumn{2}{|c|}{BP3}\\\hline\hline
\multicolumn{3}{||c||}{8 TeV} & \multicolumn{2}{|c|}{0.945} & \multicolumn{2}{|c|}{0.254} & \multicolumn{2}{|c|}{0.087}\\
\multicolumn{3}{||c||}{13 TeV} & \multicolumn{2}{|c|}{10.8} & \multicolumn{2}{|c|}{3.31} & \multicolumn{2}{|c|}{1.53}
\end{tabular}
\end{ruledtabular}
 \end{table}

\begin{figure}[t]
 \includegraphics[angle=0,width=0.5\textwidth]{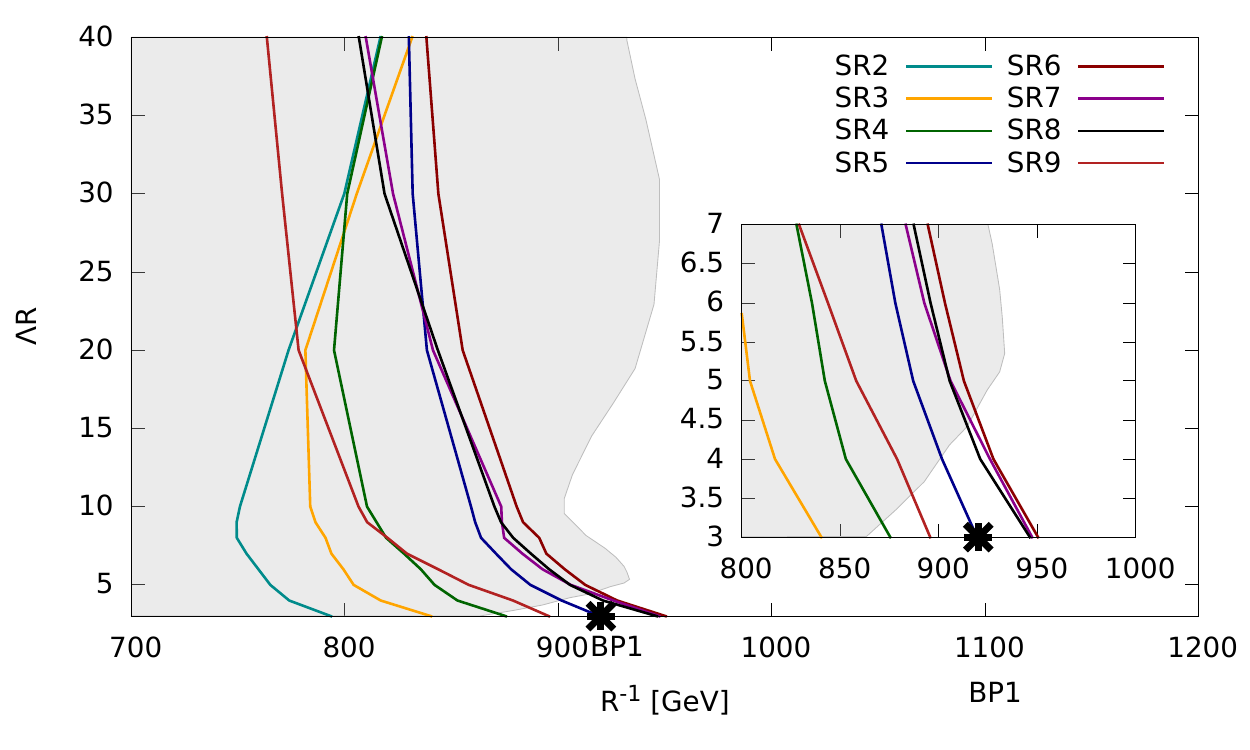}
 \caption{95\% exclusion bounds on $R^{-1}$--$\Lambda R$ plane from
   different SRs of ATLAS 8 TeV 20.3 fb$^{-1}$ integrated luminosity
   monojet-like selection criteria. Low $\Lambda R$ region is
   magnified in the inset. The shaded region corresponds to present
   ATLAS bound from dilepton plus $\cancel E_T$ search
   \cite{ATLAS_UED} at 8 TeV center-of-mass energy.}
\label{monojet_bound}
 \end{figure}

The level-1 KK-quarks (both the singlet $q_1$ and doublet $Q_1$) and
gluons ($g_1$) are copiously produced in pairs at the LHC. These,
subsequently, decay into SM particles and $B_1$ via cascades
involving other level-1 KK-particles.  As the spectra in
Table~\ref{BP} suggest, the $g_1$ can decay to both singlet ($q_1$)
and doublet ($Q_1$) quarks with almost same branching ratios, with
only a slight kinematic preference for the former. The singlet quark
can decay only to $B_1$ and SM quark. On the other hand, the
doublet quarks decay mostly to $W_1$ or $Z_1$.  Hadronic decay modes
of the $W_1$ being closed kinematically, it decays universally to all
level-1 doublet lepton flavours ($L_1$ or $\nu_1$), namely, $W_1 \to
L_{i1}^{\pm} \nu_{i0}$ and $W_1 \to L_{i0}^{\pm} \nu_{i1}$ have equal
branching ratios. Similarly, $Z_1$ can decay only to $L_1 l$ or $\nu_1
{\nu}$ (with branching fractions being determined by the corresponding
SM couplings).  The KK leptons finally decay to the invisible
$B_1$ and a ordinary (SM) lepton. Therefore, pair production of KK-quarks and/or
gluon gives rise to jets + leptons + missing transverse energy
$\cancel E_T$ signatures in the LHC experiments.  However, due to the
small splitting of level-1 KK-masses, in signal events, the
jets/leptons as well as $\cancel E_T$ are, in general, soft thereby
rendering the task a challenging one.  Collider phenomenology of
leptonic final states of mUED was discussed in Ref.~\cite{BK} with a
conclusion of opposite sign dilepton channel being the most promising
for moderate $10R^{-1} <\Lambda < 40 R^{-1}$. In order to enhance the
signal to background ratio for soft signal leptons, the idea of
imposing upper bounds on the lepton transverse momenta as well as on
the invariant masses of lepton pairs was proposed in
Ref.~\cite{DK}. Recently, the ATLAS collaboration, using 20.3
fb$^{-1}$ integrated luminosity data for proton-proton collisions at
$\sqrt s=8$ TeV has performed a dedicated search for soft dimuons
\cite{ATLAS_UED} (characterized by $6~{\rm GeV}< p_T^{\rm
  muon}<25~{\rm GeV}$ and invariant mass cuts) specially designed to
probe the mUED parameter space.  In the absence of any significant
excess of signal events over the SM backgrounds, they exclude, at 95\%
CL, the part of the parameter space, viz. the $(R^{-1}, \Lambda R)$
plane, depicted in grey in Fig.~\ref{monojet_bound}.  Clearly, for
large $\Lambda R \sim 35$, any $R^{-1}$ below about 950 GeV is ruled
out. However, for small $\Lambda R \sim 3$ (which is particularly
motivated from the stability of scalar potential with a 126 GeV Higgs
boson), the lower bound on $R^{-1}$ is only about 860 GeV, a
consequence of the very small splitting between level-1 KK-particles
and, consequently, soft leptons evading the acceptance cuts.  Hence,
an alternative search strategy is called for.

 \begin{table}
 \caption{Definitions of SRs for monojet-like selection used by ATLAS collaboration in Ref.~\cite{ATLAS8_monojet} along with 95\% CL upper limits ($\langle \epsilon \sigma \rangle_{\rm obs}^{95}$) on the product of cross section, acceptance and efficiency ($\sigma \times A \times \epsilon$). mUED cross-sections for the BPs are also presented.}
\label{monojet_selection}
 \begin{ruledtabular}
\begin{tabular}{||c||c|c||c||c|c|c||}
\multicolumn{7}{||c||}{Monojet-like Selection criteria}\\\hline\hline
Preselection & SRs & $E_T\!\!\!\!\!/~>$ & $\langle \epsilon \sigma \rangle_{\rm obs}^{95}$ & \multicolumn{3}{|c||}{$\sigma({\rm mUED})$ [fb]}\\\cline{4-7}

        &    & [GeV] & [fb] & BP1 & BP2 & BP3 \\\hline\hline
$\cancel E_T>$ 150 GeV & SR1 & 150 & 726& 70& 15& 28\\
At least one jet              & SR2 & 200 &194 &52 &11 &21\\       
with $p_T>$ 30 GeV            & SR3 & 250 &90 & 37&7.6 &13\\       
$|\eta|<$ 4.5                 & SR4 & 300 &45 & 26& 5.2&7.4\\       
Lepton veto                   & SR5 & 350 &21 & 19& 3.8&3.6\\\cline{1-1}
Monojet-like Sec.             & SR6 & 400 &12 & {\bf 14}& 2.8&1.6\\\cline{1-1}
$p_T^{j_1}>120$ GeV             & SR7 & 500 &7.2 & {\bf 7.6}& 1.5&0.53\\       
 $|\eta^{j_1}|<2.0$, $\frac{p_T^{j_1}}{\cancel E_T}>0.5$ & SR8 & 600 &3.8 &{\bf 4.0} &0.82 &0.26\\       
$\Delta \phi({\rm jet},\vec {\cancel E}_T)>1.0$& SR9 & 700 &3.4 & 2.2 & 0.46&0.13\\      

 \end{tabular}
 \end{ruledtabular}
 \end{table}

A final state comprising a single jet, resulting primarily from
initial state radiation, accompanied by a missing transverse energy
could be a promising channel. Indeed, monojet plus $\cancel E_T$ is a
very effective channel for theories with a quasi-degenerate spectrum,
for example, in the search for third-generation squarks in compressed
supersymmetry scenarios \cite{Aad:2014nra}. The spectrum of ISR jets
depends on the scale and dynamics of the production process and is
independent of the subsequent decay, including mass splittings.  With
the system of the pair of KK-particles ($g_1 g_1, g_1 q_{i1}, q_{i1}
q_{j1}, q_{i1} \bar q_{j1}$) recoiling against a hard ISR jet, the
final state comprises of a hard jet, substantial $\cancel E_T$ and
some soft jets/leptons that may or may not be visible.  Analyzing 20.3
fb$^{-1}$ data from the 8 TeV run, the CMS~\cite{CMS8_monojet} and
ATLAS~\cite{ATLAS8_monojet} collaborations have used this channel to
look for signatures of compressed SUSY, a generic DM candidate, large
extra dimensions, very light gravitinos in a gauge-mediated
supersymmetric model etc..  In view of the consistency of experimental
data and SM background predictions, model independent upper limits are
set on the product of cross section, acceptance and efficiency
($\sigma \times A \times \epsilon$).  Using the ATLAS
results~\cite{ATLAS8_monojet}, we now perform an analogous exercise
for mUED.

We generate the parton level events corresponding to pair
production of KK-quarks/gluons using the mUED implementation
\cite{ElKacimi:2009zj} for the
event generator PYTHIA \cite{pythia}. We
use the CTEQ6L1 \cite{Pumplin:2002vw} parton distributions 
with the factorization and renormalization scales kept fixed at 
the parton
center-of-mass energy. ISR, decay of KK-quarks/gluons, showering and
hadronization are also simulated with PYTHIA.  For the reconstruction of 
physics objects
(jets, leptons, $\cancel E_T$ etc.)  we closely
follow the prescription of Ref.~\cite{ATLAS8_monojet} for ATLAS
monojet + $\cancel E_T$ analysis. 
Jet candidates are
reconstructed using FastJet \cite{fjet} with the 
anti--$k_T$ jet
clustering algorithm~\cite{antikt} with a distance parameter of
0.4. Only jets with $p_T>30$ GeV and $|\eta|<4.5$ are considered for
further analysis. Electron (muon) candidates are required to have $p_T
> 7$ GeV and $|\eta| < 2.47(2.5)$. After identifying jets and leptons,
overlaps between identified electrons and jets in the final state are
resolved by discarding any jet candidate lying within a distance
$\Delta R = \sqrt{\Delta \eta^2 + \Delta \phi^2} < 0.2$ of an electron
candidate. Missing transverse momentum is reconstructed using all 
remaining visible entities, viz. jets, 
leptons and all calorimeter clusters not associated
to such objects.  After object reconstruction, only 
events with zero
lepton, $\cancel E_T > 150$ GeV 
and atleast one jet (satisfying the aforementioned 
preselection criteria) are selected for further analysis. 

A monojet-like final state topology is demanding a leading jet with
$p_T>120$ GeV, $|\eta|<2.0$ and $p_T/\cancel E_T > 0.5$. An additional
requirement on the azimuthal separation $\Delta \phi(jet,\vec{\cancel
  E_T}) > 1.0$ between the direction of the missing transverse
momentum and that of each of the selected jets is also imposed. After
selecting events with monojet-like topology, different signal regions
(SR1--SR9) are defined with progressively increasing thresholds for
$\cancel E_T$. the ATLAS monojet-like selection criteria and signal
regions are summarized in Table~\ref{monojet_selection}. For each of
these signal regions, the good agreement between the numbers of events
observed by the ATLAS detector and expected within the SM can be used
to impose model-independent upper limits on the product $\sigma \times
A \times \epsilon$, and these too are presented in
Table~\ref{monojet_selection}. These should be compared with the
monojet cross-sections for the mUED BPs which are presented in the last
three columns of Table~\ref{monojet_selection}. Clearly, for BP1, the
signal cross-sections exceed the ATLAS 95\% CL upper limits ($\langle
\epsilon \sigma \rangle_{\rm obs}^{95}$) for each of SR6, SR7 and
SR8. Thus, BP1, which had survived the dimuon search bounds ($R^{-1} =
860$ GeV for $\Lambda R \sim 3$), is squarely ruled out by the monojet
analysis. While this may seem only a modest improvement, given that
each of BP2 and BP3 survive, as we shall see below, this is crucial
for a hole in the parameter space would have been left
otherwise. Indeed, for low $\Lambda R$ (preferred in the context of
the stability of the Higgs potential), this constitutes the most
promising channel.  Our final exclusion limits in the $R^{-1}-\Lambda
R$ plane from different SRs of ATLAS $\sqrt s=$8 TeV and ${\cal
  L}=$20.3 fb$^{-1}$ monojet+ $\cancel E_T~$ analysis are presented in
Fig.~\ref{monojet_bound}.

 \begin{table}[t]
 \caption{Definition of SRs for multijets plus $\cancel E_T$
   analysis used by ATLAS collaboration in
   Ref.~\cite{ATLAS13_multijets} for 13 TeV center-of-mass energy and
   3.2 inverse femtobern integrated luminosity.  
    $\Delta \phi(j,\vec{\cancel E}_T)$ is the azimuthal separations 
   between $\vec{\cancel E}_T$ and the reconstructed jets. ${m_{eff}(N_j)}$ is
   defined to be the scalar sum of the transverse momenta of the
   leading $N$ jets together with $\cancel E_T~$. However, for
   $m_{eff}^{incl.}$, the sum goes over all jets with $p_T>50$
   GeV. Model independent 95\% CL upper limits on multijets $\langle
   \epsilon \sigma \rangle_{\rm obs}^{95}=\sigma \times A \times
   \epsilon$ and mUED cross-sections for the BPs are also presented.}
\label{multijets_selection}
 \begin{ruledtabular}
\begin{tabular}{||c||c|c|c||c||c||c|c||}
Cuts & \multicolumn{7}{|c||}{Signal Region}\\\cline{2-8}
  & 2jL & 2jM & 2jT & 4jT & 5j & 6jM & 6jT \\\hline\hline
$\cancel E_T~>$ [GeV] &200 &200 &200 &200 &200 &200 &200 \\
$p_T^{j_1}>$ [GeV] &200 &300 &200 &200 &200 &200 &200 \\
$p_T^{j_2}>$ [GeV] &200 &50 &200 &100 &100 &100 &100 \\
$p_T^{j_3}>$ [GeV] &- &- &- &100 &100 &100 &100 \\
$p_T^{j_4}>$ [GeV] & -&- &- &- &100 & 100&100 \\
$p_T^{j_5}>$ [GeV] & -& -& -& -& -& 100&100 \\
$p_T^{j_6}>$ [GeV] &- &- &- &- &- & 100 &100 \\
$\Delta \phi(j_{<3},\vec {\cancel E}_T)>$ &0.8 &0.4 &0.8 &0.4 &0.4 &0.4 &0.4 \\
$\Delta \phi(j_{>3},\vec {\cancel E}_T)>$ & -&- &- &0.2 &0.2 &0.2 &0.2 \\ 
$\frac{\cancel E_T~}{\sqrt H_T}~[\sqrt{\rm GeV}]>$ &15 &15 &20 &- &- &- &- \\
Aplanarity$>$ &- & -& -&0.04 &0.04 &0.04 &0.04 \\
$\frac{\cancel E_T~}{m_{eff}^{N_j}}>~[\sqrt{\rm GeV}]$ &- &- &- &0.2 &0.25 &0.25 &0.2 \\
$m_{eff}^{\rm incl.}>$ [GeV] &1200 &1600 &2000 &2200 &1600 &1600 &2000 \\\hline\hline
$\langle \epsilon \sigma \rangle_{\rm obs}^{95}$ [fb] & 24 & 21 & 5.9 & 2.5 & 2.0 & 1.6 & 1.6 \\\hline\hline
BP1 & {\bf 36}& {\bf 66}& {\bf 11}& 2.2& 0.97& 0.22& 0.11\\
BP2 & 11& {\bf 24}& 3.6& 0.93& 0.5& 0.13&0.13 \\
BP3 & 13& 14& 2.8& 2.3& {\bf 2.2}& 0.46& 0.29\\
 \end{tabular}
 \end{ruledtabular}
 \end{table}

The situation can be further improved if other channels are considered
as well. A particularly useful one is that constituting multijets +
$\cancel E_T$.  Recently, both the ATLAS \cite{ATLAS13_multijets} and
the CMS \cite{CMS13_multijets} collaborations have communicated
results for such an analysis for $\sqrt s=$13 TeV, although for only a
small data set (3.2 fb$^{-1}$).  This, however, can be offset by an
increased cross section. At the LHC, the dominant contribution to the
pair production of level-1 KK-quarks/gluons arises from gluon-gluon or
quark-gluon initial states.  The gluon density increases by an order
of magnitude as we go from 8 TeV to 13 TeV.  Similarly, the presence
of $t$-channel diagrams as well as the momentum-dependence of the
vertices largely compensates for any suppression of the leading
parton-level cross sections with an increase in the subprocess center
of mass energy.  For example, as Table~\ref{BP} shows the total mUED
production cross-section increases by a factor about 20 for
KK-quark/gluon mass $\sim$ 1400 GeV. Therefore, it is instructive to
study 13 TeV multijets + $\cancel E_T~$ results in the context of mUED
which we will discuss in the following.

 \begin{figure}[h]
 \includegraphics[angle=0,width=0.5\textwidth]{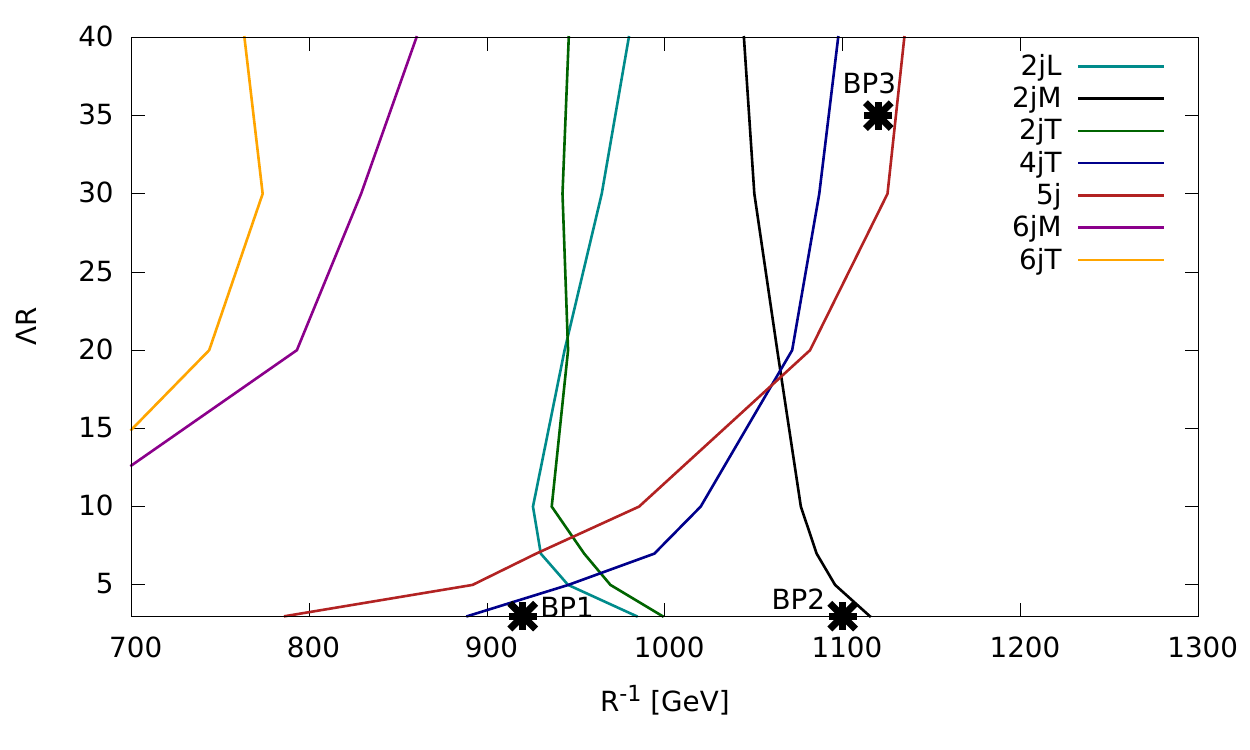}
 \caption{95\% exclusion bounds on $R^{-1}$--$\Lambda R$ plane from different SRs of ATLAS 13 TeV 3.2 fb$^{-1}$ integrated luminosity multijets plus $\cancel E_T~$ analysis.}
\label{multijets_bound}
 \end{figure}

The aforementioned ATLAS analysis\cite{ATLAS13_multijets} searched for
events with 2--6 jets in association with a large $\cancel E_T~$.  Jet
candidates are reconstructed using the anti-kt jet clustering
algorithm with a distance parameter of 0.4, and only those with $p_T >
20$ GeV and $|\eta|<2.8$ are retained. Electron (muon) candidates are
required to have $p_T > 10$ GeV and $|\eta| < 2.47(2.5)$. Furthermore,
a lepton needs to be isolated
from a  jet for the two entities to be reconstructed unambiguously. 
Consequently, any putative jet falling within an
angular distance $\Delta R=0.2$ of 
an electron is not reconstructed into a jet, leaving the 
constituents as unattached objects at that stage. 
Similarly, any electron falling within $\Delta R=0.4$ of a surviving jet 
candidate is not considered as one, and its energy-momentum ascribed 
to the jet. 
The missing transverse momentum is
reconstructed using all the remaining jets and leptons as well as all
calorimeter clusters not associated to such objects. After the object
reconstruction, events containing lepton(s) with $p_T>10$ GeV are
vetoed. Jets with $p_T>50$ GeV are considered for further analysis.

The ATLAS collaboration considers seven inclusive analysis channels,
characterized by increasing jet multiplicity and different cuts to
reduce the SM background. The effective mass, $m_{\rm eff}$, and
$\cancel E_T$ turn out to be the most powerful discriminants between
the multijets signal and SM backgrounds. These additional selection
cuts are imposed on
\begin{itemize}
\item $\cancel E_T$ 
\vspace*{-7pt}
\item $m_{\rm eff}^{\rm incl.}$ defined as the scalar sum of
  $\cancel E_T$ and the $p_T$s  of all jets with
  $p_T>50$ GeV,  
\vspace*{-7pt}
\item $\cancel E_T/m_{\rm eff}^{N_j}$ (for events with {\em at
    least} $N_j$ jets) where $m_{\rm
  eff}^{N_j}=\sum_{i=1}^{N_j}p_T^i({\rm jet})+\cancel E_T~$ with
  the sum extending over the leading $N_j$ jets,
\vspace*{-7pt}
\item $\cancel E_T/\sqrt H_T$ where $H_T = m_{\rm eff}^{\rm incl.} -
  \cancel E_T$, 
\vspace*{-7pt}
\item $\Delta\phi({\rm jet}, \vec{\cancel E}_T)$, 
\vspace*{-7pt}
\item the aplanarity variable $A$ defined as $A = 3\lambda_3/2$, where
  $\lambda_3$ is the smallest eigenvalue of the normalized momentum
  tensor of the jets.
\end{itemize}
In Table~\ref{multijets_selection}, we list the cuts used by the ATLAS
collaboration to define the signal regions, and the corresponding
model independent 95\% CL upper limits on $\langle \epsilon \sigma
\rangle_{\rm obs}^{95}=\sigma \times A \times \epsilon$.

For simulating the production of level-1 KK-quarks/gluons, their
subsequent decays, ISR, showring and hadronization, we follow the same
procedure as outlined before.  The physics objects are reconstructed
and events selected to mimic the ATLAS criteria described above. The
signal cross-sections for the three BPs are presented in the last
three rows of Table~\ref{multijets_selection}. BP1 and BP2 being
characterized by low $\Lambda R$ (and, hence, small splittings),
typically give rise to low jet multiplicities. Whereas BP2 is seen to
be ruled out by the `2jT' criteria, BP1 is ruled out by each of the
three dijet SRs. On the other hand, BP3, owing to the larger $R^{-1}$,
is associated with a smaller total cross section and easily evades the
dijet constraints. However, owing to the much larger $\Lambda R$, the
relative splittings are larger and a substantial fraction of events
lead to multijet configurations. This, for example, allows us to rule
it out using the `5j' SR.

In Fig.~\ref{multijets_bound}, we present the final exclusion bounds
(drawn from the ATLAS analysis of the 3.2 fb$^{-1}$ data collected in
the 13 TeV run) in the mUED parameter space for each of the SRs listed
in Table~\ref{multijets_selection}. The region in the $(R^{-1},
\Lambda R)$ to the right of a given curve is ruled out at 95\%
C.L. Note that for large $\Lambda R \, (\gapp 30)$, the strongest
bounds come from an analysis of final state with at least 5 jets and
is about 1130 GeV. The sensitivity falls drastically for the inclusive
six-jet final state. This can be understood by realizing that, at the
parton level, the decay of the KK-particles would lead to at most four
SM quarks/gluons (and, that too only for $g_1 g_1$ production).  For
low $\Lambda R \, (\lapp 16)$, on the other hand, the strongest bound
($R^{-1} \gapp 1110$ GeV) is achievable from the `2jM' signal
region. It is worthwhile to note that the basic requirements (one jet
with $p_T>300$ GeV, another jet with $p_T>50$ GeV and $\cancel
E_T~>300$ GeV) for the `2jM' signal region is markedly similar to the
monojet-like selection criteria used earlier. Hence, it comes as no
surprise that this constitutes the most efficient strategy for low
$\Lambda R$.

 \begin{figure}[h]
 \includegraphics[angle=0,width=0.5\textwidth]{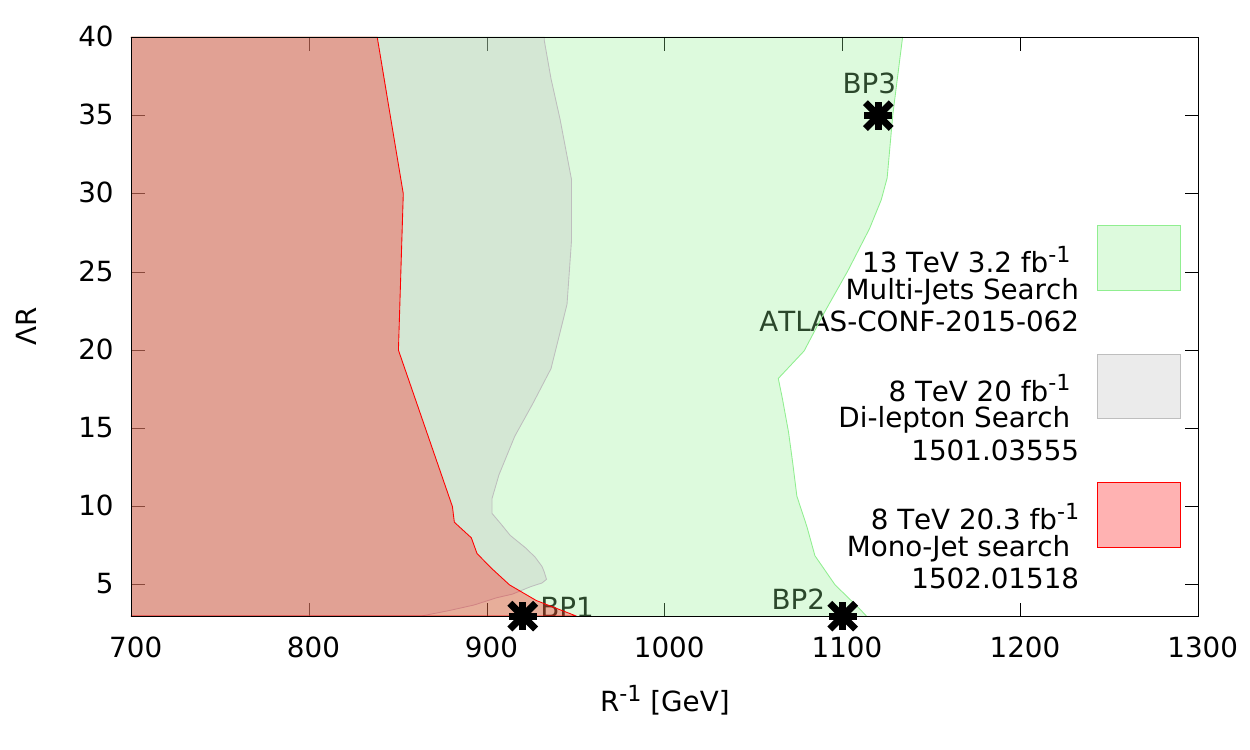}
 \caption{Regions of $R^{-1}$--$\Lambda R$ plane excluded from ATLAS Run I monojet and lepton \cite{ATLAS_UED} analysis as well as Run II 3.2 fb$^{-1}$ multijets plus $\cancel E_T~$ searches. }
\label{bound}
 \end{figure}

To summarize, we have computed constraints on the mUED parameter space
using the ATLAS 8 TeV monojet results. While the dedicated ATLAS
search strategy (involving soft dimuons) is a promising one for large
$\Lambda R$, it is not very efficient for low $\Lambda R$, a region
preferred by certain theoretical considerations, such as the stability
of the Higgs potential. On the other hand, the monojet channel, being
independent of the mass splitting (and, hence, $\Lambda R$) does not
suffer from the drawbacks of the soft dimuon channel, and is seen to
be much more sensitive (in this region) than the latter.  This,
clearly, calls for the inclusion of mUED as a candidate scenario for
any future monojet study at ATLAS/CMS.  We also examine the efficacy
of the multijet ($+ \cancel E_T$) signal in the context of mUED. Even
with the small sample size analyzed by the ATLAS collaboration, this
is shown to lead to a much stronger exclusion of the mUED parameter
space, as is depicted in Fig.~\ref{bound}, wherein the different
shaded regions correspond to the individual exclusions allowed by
different search strategies.  A casual perusal of Fig.~\ref{bound}
might suggest that the multijet channel is overwhelmingly
superior. This should be treated with caution, though.  For one, the
particular sub-channel that is the most sensitive one for $\Lambda R
\lapp 16$ is the one that is remarkably close to the monojet
algorithm, experimental results for which do not yet exist for the 13
TeV run. Similarly, the inclusion of multijet final state alongwith
soft but isolated leptons (i.e., without the lepton veto being imposed
as was done in the ATLAS analysis) is likely to lead to some
improvement in the sensitivity.

DC acknowledges partial support from the European Union's Horizon 2020
research and innovation programme under Marie Sklodowska-Curie grant
No 674896, and the R\&D grant of the University of Delhi. KG is
supported by DST (India) under INSPIRE Faculty Award.

\end{document}